\documentclass[aps,titlepage,11pt,superscriptaddress]{revtex4-1}
\usepackage[dvips]{graphicx}
\usepackage{times}
\usepackage{color}
\usepackage{amssymb, amsmath}
\usepackage{lineno}

\begin{document}

\title{\Large\textbf{Regulation of Irregular Neuronal Firing by Autaptic Transmission}}

\author{Daqing Guo}
\thanks{Electronic address: \textcolor{blue}{dqguo@uestc.edu.cn}}
\affiliation{Key Laboratory for NeuroInformation of Ministry of Education, School of Life Science and Technology, University of Electronic Science and Technology of China, Chengdu 610054, People's Republic of China}
\affiliation{Center for Information in Medicine, University of Electronic Science and Technology of China, Chengdu 610054, People's Republic of China}

\author{Shengdun Wu}
\affiliation{Key Laboratory for NeuroInformation of Ministry of Education, School of Life Science and Technology, University of Electronic Science and Technology of China, Chengdu 610054, People's Republic of China}

\author{Mingming Chen}
\affiliation{Key Laboratory for NeuroInformation of Ministry of Education, School of Life Science and Technology, University of Electronic Science and Technology of China, Chengdu 610054, People's Republic of China}

\author{Matja{\v z} Perc}
\affiliation{Department of Physics, Faculty of Natural Sciences and Mathematics, University of Maribor, Maribor, Slovenia}

\author{Yangsong Zhang}
\affiliation{School of Computer Science and Technology, Southwest University of Science and Technology, Mianyang, China}

\author{Jingling Ma}
\affiliation{Key Laboratory for NeuroInformation of Ministry of Education, School of Life Science and Technology, University of Electronic Science and Technology of China, Chengdu 610054, People's Republic of China}

\author{Yan Cui}
\affiliation{Key Laboratory for NeuroInformation of Ministry of Education, School of Life Science and Technology, University of Electronic Science and Technology of China, Chengdu 610054, People's Republic of China}

\author{Peng Xu}
\affiliation{Key Laboratory for NeuroInformation of Ministry of Education, School of Life Science and Technology, University of Electronic Science and Technology of China, Chengdu 610054, People's Republic of China}
\affiliation{Center for Information in Medicine, University of Electronic Science and Technology of China, Chengdu 610054, People's Republic of China}

\author{Yang Xia}
\affiliation{Key Laboratory for NeuroInformation of Ministry of Education, School of Life Science and Technology, University of Electronic Science and Technology of China, Chengdu 610054, People's Republic of China}
\affiliation{Center for Information in Medicine, University of Electronic Science and Technology of China, Chengdu 610054, People's Republic of China}

\author{Dezhong Yao}
\affiliation{Key Laboratory for NeuroInformation of Ministry of Education, School of Life Science and Technology, University of Electronic Science and Technology of China, Chengdu 610054, People's Republic of China}
\affiliation{Center for Information in Medicine, University of Electronic Science and Technology of China, Chengdu 610054, People's Republic of China}

\begin{abstract}
The importance of self-feedback autaptic transmission in modulating spike-time irregularity is still poorly understood. By using a biophysical model that incorporates autaptic coupling, we here show that self-innervation of neurons participates in the modulation of irregular neuronal firing, primarily by regulating the occurrence frequency of burst firing. In particular, we find that both excitatory and electrical autapses increase the occurrence of burst firing, thus reducing neuronal firing regularity. In contrast, inhibitory autapses suppress burst firing and therefore tend to improve the regularity of neuronal firing. Importantly, we show that these findings are independent of the firing properties of individual neurons, and as such can be observed for neurons operating in different modes. Our results provide an insightful mechanistic understanding of how different types of autapses shape irregular firing at the single-neuron level, and they highlight the functional importance of autaptic self-innervation in taming and modulating neurodynamics.
\end{abstract}

\maketitle

\section*{Introduction}
Without doubt cortical neurons operate in noisy environments \cite{DestexheandRudolph-Lilith2012}. There is a
broad consensus that neuronal noise can exert a strong impact on stochastic dynamics of neurons \cite{BalenzuelaandGarcia-Ojalvo2006, Zaikinetal2003, ZhouandKurths2003} and drive them to discharge action potentials or so-called ``spikes'' highly irregular \cite{StevensandZador1998}. In animal experiments, temporally irregular firing of neurons has been widely observed, both during ongoing spontaneous activity and when driven at high firing rates \cite{Stiefeletal2013,Fellousetal2003, Destexheetal2003}. Importantly, irregular neuronal firing has been linked to several higher brain cognitive functions, such as working memory \cite{HanselandMato2013}, selective attention \cite{Ardidetal2010} and sensory coding \cite{Doronetal2014}. Many theoretical studies have shown that the degree of neuronal firing irregularity might be non-monotonic dependent on neuronal noise intensity \cite{GuoandLi2009, PikovskyandKurths1997, Ozeretal2009}. At an optimal level of neuronal noise, neurons may exhibit coherent firing of spikes, indicating the occurrence of counterintuitive phenomenon termed as ``coherence resonance'' (CR) \cite{Kreuzetal2006, Lucciolietal2006}. Furthermore, CR has been observed in several biological experiments and has been proposed as one basic mechanism that neurons may use to facilitate signal transmission \cite{Manjarrezetal2002}.

By weighting and combining outside signals, neurons continuously emit sequences of action potentials of their own. Past experimental evidence indicated that neurons receive roughly equal average amount of depolarizing and hyperpolarizing currents from their presynaptic neurons \cite{Haideretal2006, Marinoetal2005, HigleyandContreras2006}. Such balanced excitation-inhibition synaptic bombardment is regarded as an important source of neuronal noise, which is essential for triggering irregular neuronal firing \cite{Destexheetal2003, Fellousetal2003, ShadlenandNewsome1998}. In the balanced excitation-inhibition state, firing irregularity of neurons highly depends on the intensity of synaptic bombardment. Under this condition, the overall effects of excitation and inhibition are nearly canceled, and thus neuronal firing is driven by fluctuations that transiently spoil this cancellation \cite{StevensandZador1998, PikovskyandKurths1997, GuoandLi2010}. Using simplified models, previous studies have demonstrated that several key intrinsic proprieties of balanced excitation-inhibition bombardment from presynaptic neurons, such as the mean firing input rate and synaptic coupling strength, contribute greatly to the regulation of irregular neuronal firing \cite{Kreuzetal2006, Lucciolietal2006}.

In addition to normal ``feedforward'' synapses, neurons also form self-feedback connections, termed as ``autapses'', onto themselves \cite{Baccietal2003, Tamasetal1997, Lubkeetal1996}. Autapses have been frequently observed in various brain regions \cite{Parketal1980, KarabelasandPurpura1980, Cobbetal1997}, and their kinetics have been identified to exhibit the similar electrical properties as normal synapses \cite{Yamaguchi2009}. Using both experimental and computational approaches, several studies have revealed that autaptic self-innervation of neurons might play functional roles in modulating neuronal dynamics. For instance, recent electrophysiological recordings have indicated that fast-spiking interneurons with GABAergic autaptic transmission have relatively higher levels of firing precision than those without GABAergic self-connections \cite{BacciandHuguenard2006}. By computational modelling, it has been found that autapses can intermittently control synchronization \cite{Wangetal2015}, mediate propagation of weak rhythmic activity \cite{Yilmazetal2016} and induce spiral wave \cite{Qinetal2015} in neuronal networks. In addition, autapses have also been postulated to improve synaptic transmission reliability \cite{Yamaguchietal1997}. Theoretically, self-feedback neuronal activities from chemical and electrical autpases might also influence the neuronal firing dynamics and thus regulate firing regularity of neurons. However, so far the precise roles of different types of autapses in shaping irregular neuronal firing are still not completely established.

In this study, we simulate a spiking model neuron driven by both the balanced excitation-inhibition presynaptic inputs from presynaptic neurons and the autaptic input from itself (see Fig.~1, and Materials and Methods). Using various measurements, we perform analysis for spiking activities generated by the model neuron under different autaptic coupling conditions. We identify that both chemical and electrical autapses participate in the modulation of neuronal irregular firing, primarily through mediating the frequency of burst firing. Further investigation confirms that our critical results are independent of single neuron firing properties and can be observed for neurons operating in different modes. These findings highlight the functional roles of autapses in the regulation of neuronal firing irregularity.

\section*{Results}
We firstly consider a single Izhikevich neuron with calss I excitability~\cite{Izhikevich2003}, and examine how purely presynaptic bombardment controls its firing irregularity. Then, we incorporate the autaptic current into the model (see Figure1) and systemically investigate the detailed roles of different types of autapses in shaping irregular firing for neurons displaying class I excitability. Finally, we further extend our results to spiking neurons with class II and III excitabilities.

\begin{figure}[!t]
\includegraphics[width=16cm]{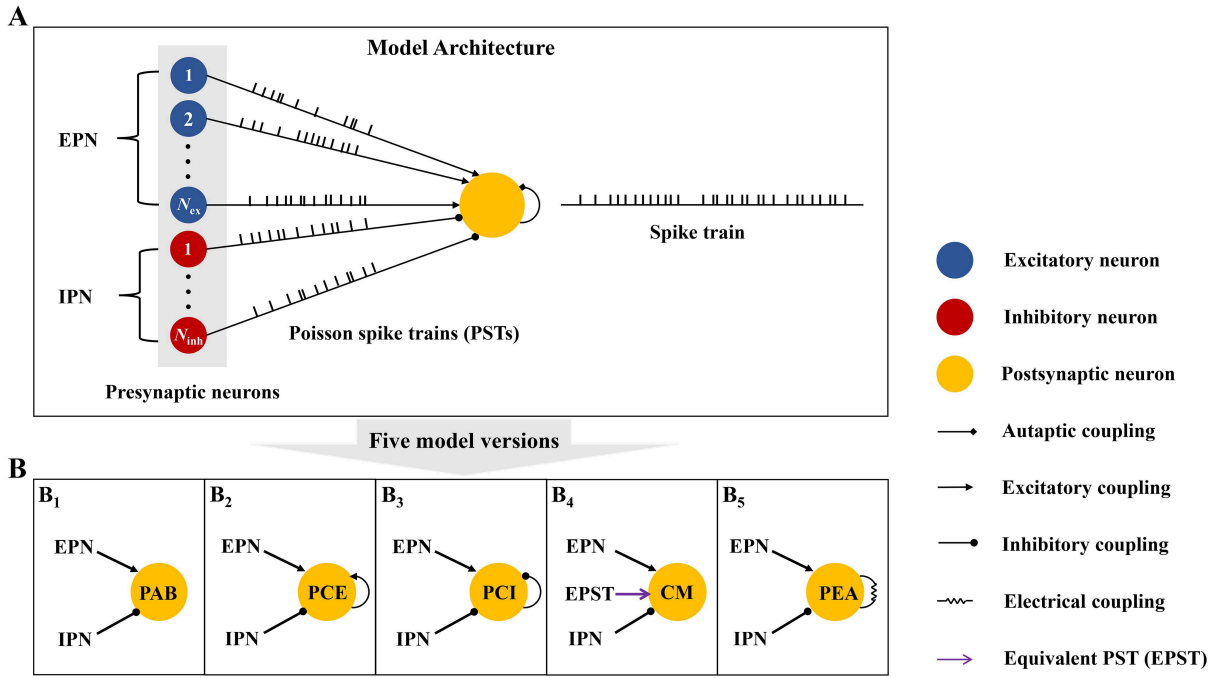}
\caption{(Color online)
Schematic description of the computational model. A:~Basic model architecture. In the model, the postsynaptic neuron receives balanced excitation-inhibition input from $N_{\text{ex}}$ excitatory presynaptic neurons (EPN) and $N_{\text{inh}}$ inhibitory presynaptic neurons (IPN). For simplicity, each presynaptic neuron is modelled as a Poisson spike train generator, with a fixed input rate $f_{\text{in}}$. In addition, the postsynaptic neuron is also driven by the self-feedback autaptic input from itself. B:~Five model versions used in our simulations. From (B$_1$)-(B$_5$), five models are termed as: the postsynaptic neuron with autapse blockade (PAB), the postsynaptic neuron with chemical excitatory autapse (PCE), the postsynaptic neuron with chemical inhibitory autapse (PCI), the comparative model (CM), and the postsynaptic neuron with electrical autapse (PEA), respectively. Note that the CM model is designed to compare with either the PCE or PCI model, in which the chemical autapse is replaced by an equivalent Poisson spike train (EPST) with both the same coupling type and firing rate.}
\label{fig:1}
\end{figure}

\subsection*{Presynaptic bombardment controls irregular neuronal firing and triggers coherence resonance}

\begin{figure}[!t]
\includegraphics[width=16cm]{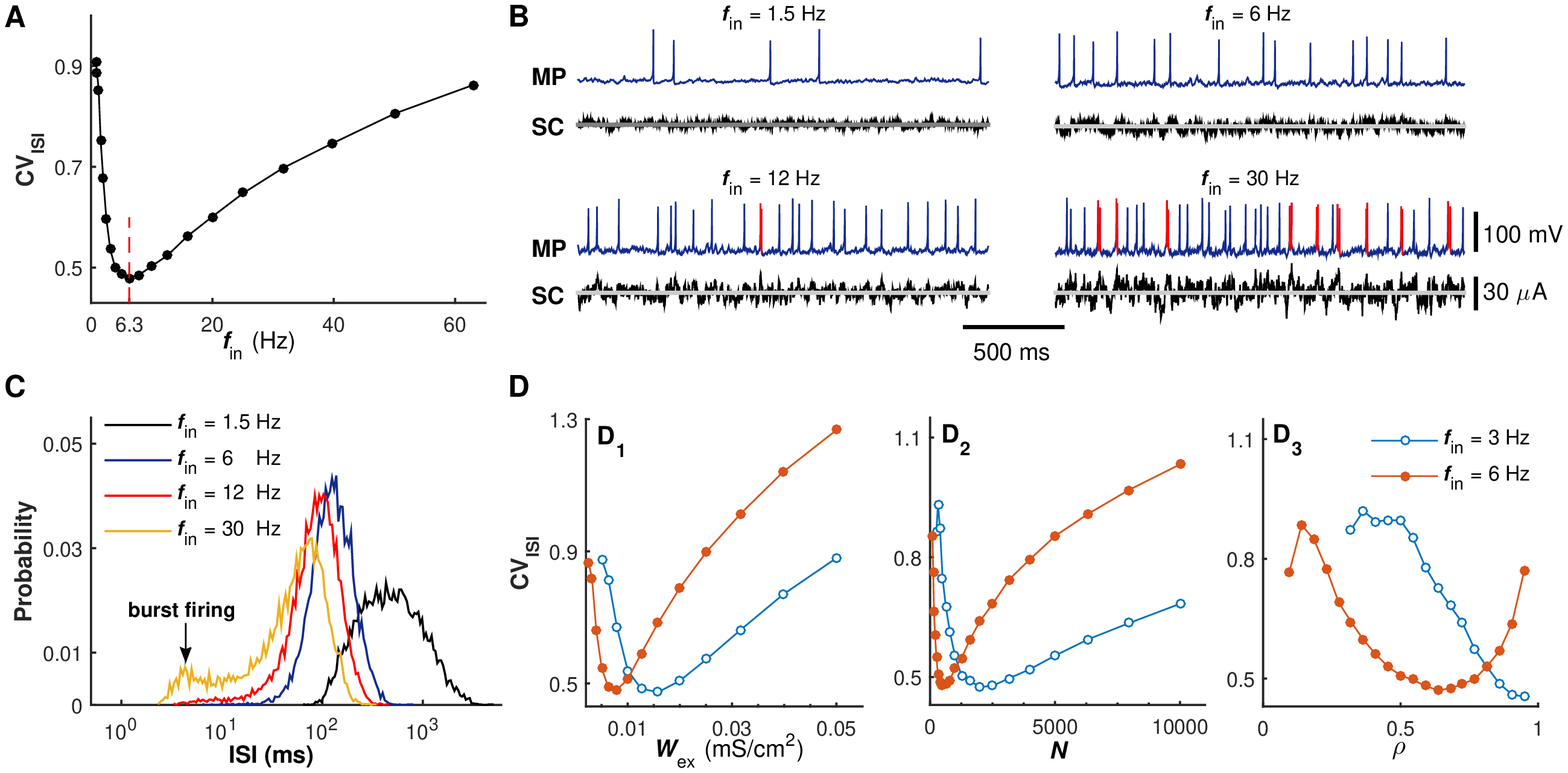}
\caption{(Color online)
Presynaptic bombardment contributes to the modulation of neuronal firing irregularity.  A:~The $\text{CV}_{\text{ISI}}$ value is plotted as a function of the input rate $f_{\text{in}}$ for the PAB model. The postsynaptic neuron achieves the best firing regularity at $f_{\text{in}}=6.3$~Hz. B:~Typical membrane potential (MP) traces and corresponding synaptic current (SC) traces at different input rates. Here the red color in MP traces denotes the occurrence of burst firing. The black lines in SC traces represent the total synaptic currents from presynaptic neurons, and the gray lines in SC traces are zero-current levels. Four input rates considered in (B) are: $f_{\text{in}}=1.5$~Hz, $f_{\text{in}}=6$~Hz, $f_{\text{in}}=12$~Hz and $f_{\text{in}}=30$~Hz. C:~ISI distribution curves correspond to the above four input rates. Each ISI distribution curve is computed using 10$^5$ firing events. D:~Dependence of the $\text{CV}_{\text{ISI}}$ value on three key presynaptic-related parameters, which are the excitatory synaptic strength (D$_1$), the population size of presynaptic neurons (D$_2$) and the proportion of excitatory neurons (D$_3$). Two input rates considered in (D) are: $f_{\text{in}}=3$~Hz and $f_{\text{in}}=6$~Hz.}
\label{fig:1}
\end{figure}

We begin by examining how the presynaptic bombardment modulates the firing regularity of the considered model neuron \cite{ShadlenandNewsome1998, Kreuzetal2006, Lucciolietal2006}. To this end, we artificially block the autaptic coupling and stimulate the postsynaptic neuron with different presynaptic input rates. For simplicity, we call the postsynaptic neuron with autapse blockade as the PAB model (Fig.~1B$_1$). Figure~2A shows the coefficient of variation of inter-spike intervals ($\text{CV}_{\text{ISI}}$) of the postsynaptic neuron as a function of the input rate for the PAB model~\cite{KochandSegev1998}. With increasing the input rate, we find that the $\text{CV}_{\text{ISI}}$ curve first drops and then rises, and the smallest $\text{CV}_{\text{ISI}}$ value is achieved at an intermediate input rate of 6.3~Hz. Such finding clearly illustrates the occurrence of coherence resonance \cite{Ozeretal2009, PikovskyandKurths1997, GuoandLi2009, Kreuzetal2006, Lucciolietal2006}. When the input rate is low, the synaptic current due to presynaptic bombardment is weak and has small fluctuations (Figs.~2B and 2C, $f_{\text{in}}=1.5$~Hz). Under this condition, the postsynaptic neuron generates few scattered spikes with low-temporal coherence. As the input firing rate grows, the fluctuations of synaptic current become larger and the postsynaptic neuron firing more frequently. For an appropriate level of presynaptic bombardment, the postsynaptic neuron exhibits a strong integration capability and fires well-separated spikes in a high rate, thus having a lower $\text{CV}_{\text{ISI}}$ value (Figs.~2B and 2C, $f_{\text{in}}=6$ and 12~Hz). However, too strong presynaptic bombardment drives the postsynaptic neuron to fire bursts occasionally (Fig.~2B, $f_{\text{in}}=30$~Hz). These bursting events cause a peak at small intervals (less than 10~ms) in the ISI distribution curve (Fig.~2C, $f_{\text{in}}=30$~Hz), and therefore result in a notable enhancement in the $\text{CV}_{\text{ISI}}$ value.

In reality, the intensity of synaptic noise due to presynaptic bombardment is not only controlled by the input rate, but may be also significantly influenced by several other presynaptic-related parameters \cite{Kreuzetal2006, Lucciolietal2006, GuoandLi2012}. We therefore perform further computational studies using the PAB model to examine possible roles of different presynaptic-related parameters in the regulation of irregular firing. The results presented in Fig.~2D demonstrate our above speculation, suggesting that three other critical presynaptic-related parameters, which are the excitatory synaptic strength, the population size of presynaptic neurons and the proportion of excitatory neurons, may contribute to the modulation of irregular neuronal firing and appropriate tuning of these parameters might also trigger the occurrence of coherence resonance. Moreover, we find that irregular neuronal firing modulations caused by these three parameters are highly dependent on the input rate (Fig.~2D). For all cases, increasing the input rate moves the $\text{CV}_{\text{ISI}}$  curves toward to the low-parameter regions. Compared to the other two parameters, the modulation of irregular neuronal firing due to the proportion of excitatory neurons is more dependent on the input rate. Specifically, the coherence resonance can be triggered by tuning this parameter only for sufficiently strong input rate (Fig.~2D$_3$).

Together, our results consistently showed that the presynaptic bombardment indeed modulates irregular neuronal firing at the single-neuron level, and the relatively coherent neuronal firing can be achieved at an optimal presynaptic bombardment level. These findings are in agreement with several previous computational modelling studies \cite{Kreuzetal2006, Lucciolietal2006}, which provide us a reference basis to further investigate underlying functional roles of different types of autapses in the modulation of irregular neuronal firing.

\subsection*{Complicated roles of chemical autapses in shaping neuronal firing irregularity}

\begin{figure}[!t]
\includegraphics[width=16cm]{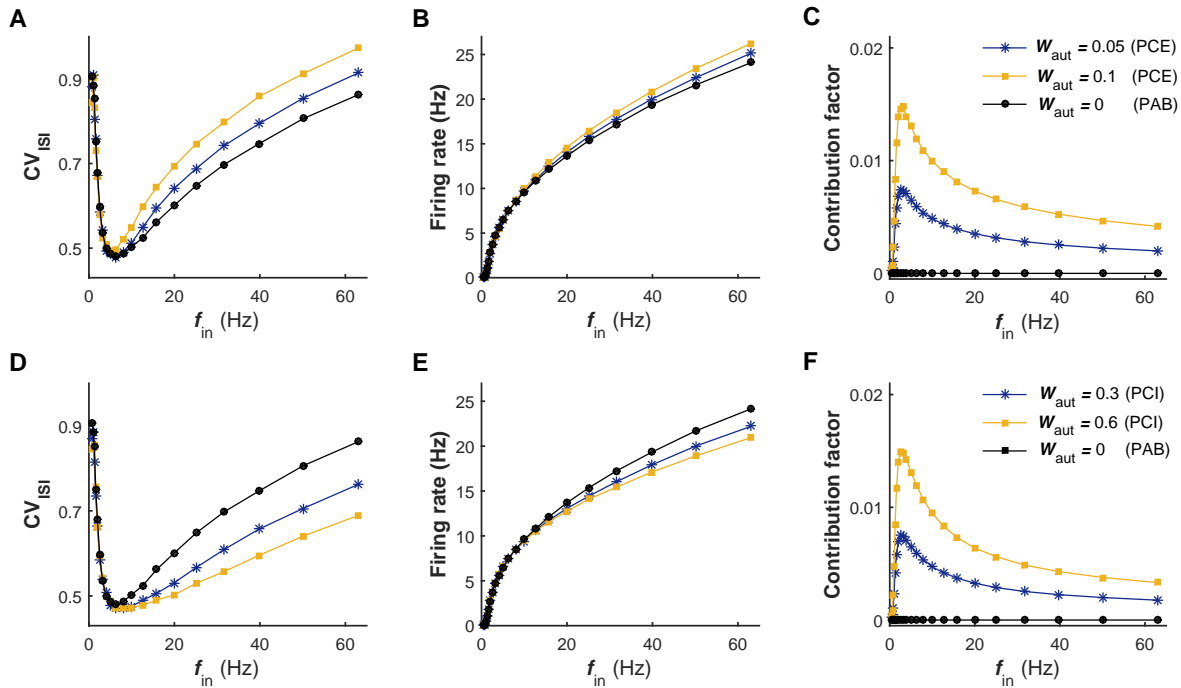}
\caption{(Color online)
Roles of chemical autapses in modulating irregular neuronal firing.  A-C: Dependence of the $\text{CV}_{\text{ISI}}$ value (A), average output firing rate (B) and autaptic contribution factor (C) on the input rate $f_{\text{in}}$ for the PCE and PAB models. In (A-C), the excitatory autaptic coupling strengths are: $W_{\text{aut}}=0.05$~mS/cm$^2$ (PCE), $W_{\text{aut}}=0.1$~mS/cm$^2$ (PCE) and $W_{\text{aut}}=0$~mS/cm$^2$ (PAB). D-F:~Dependence of the $\text{CV}_{\text{ISI}}$ value (D), average output firing rate (E) and autaptic contribution factor (F) on the input rate $f_{\text{in}}$ for the PCI and PAB models. In (D-F), the inhibitory autaptic coupling strengths are: $W_{\text{aut}}=0.3$~mS/cm$^2$ (PCI), $W_{\text{aut}}=0.6$~mS/cm$^2$ (PCI) and $W_{\text{aut}}=0$~mS/cm$^2$ (PAB). Simulations indicate that both excitatory and inhibitory autapses modulate the firing irregularity of neurons in the intermediate and high input regimes.}
\label{fig:1}
\end{figure}

To establish whether and, if possible, how chemical autapses shape neuronal firing irregularity, we open the self-feedback connection and implement two types of models: the postsynaptic neuron with excitatory autapse (the PCE model, Fig.~1B$_2$) and the postsynaptic neuron with inhibitory autapse (the PCI model, Fig.~1B$_3$). In Figs.~3A and 3D, we plot the $\text{CV}_{\text{ISI}}$ value of the postsynaptic neuron generated by these two models as a function of the input rate under different self-feedback levels, respectively. For both the PCE and PCI models, the pronounced regulations of irregular neuronal firing are observed in the intermediate and high input regimes (Figs.~3A and 3D). By comparing with the results of PAB model (i.e., $W_{\text{aut}}=0$~mS/cm$^2$), we find that the regulations of irregular neuronal firing induced by excitatory and inhibitory autapses exhibit different features (see Figs.~3A and 3D). Specifically, our results show that the autaptic excitation deteriorates neuronal firing regularity, thus increasing the $\text{CV}_{\text{ISI}}$ value. In contrast, the autaptic inhibition improves the firing regularity and results in a significant reduction in the $\text{CV}_{\text{ISI}}$ value. For both cases, these two types of regulations are found to become stronger with increasing the self-feedback level (Figs.~3A and 3D). Consequently, the postsynaptic neuron in the PCI model shows a more regular neuronal firing during intermediate and high levels of presynaptic bombardment, yielding a better CR performance (compared the results in Figs.~3A and 3D).

Theoretically, the notable modulation effects of chemical autapses on irregular neuronal firing must be due to the extra self-feedback input from the postsynaptic neuron itself. To quantify the contribution of self-feedback input to the total synaptic current, we record the average output firing rate of postsynaptic neuron and compute the contribution factor (see Materials and Methods) versus the input rate for models under different chemical autaptic coupling conditions. For both the PCE and PCI models, we find that the average firing rate of postsynaptic neuron is progressively increased with the growth of input rate (Figs.~3B and 3E). Nevertheless, due to the nonlinear input-output relation, the contribution factor curves for both excitatory and inhibitory autapses are highly nonlinear and exhibit typical bell-shaped profile (Figs.~3C and 3F). Contrary to our prediction, we surprisingly observe that for all considered cases the obtained contribution factor maintains at a low level (Figs.~3C and 3F), indicating that such self-feedback input only accounts for a small fraction of the total synaptic current (also see Fig.~4A). This counterintuitive finding reveals that the self-feedback input deriving from a chemical autapse contributes limited but, rather efficiently, to the total synaptic current, and is sufficient to trigger pronounced irregular firing regulation in the intermediate and high input regimes.

\begin{figure}[!t]
\includegraphics[width=16cm]{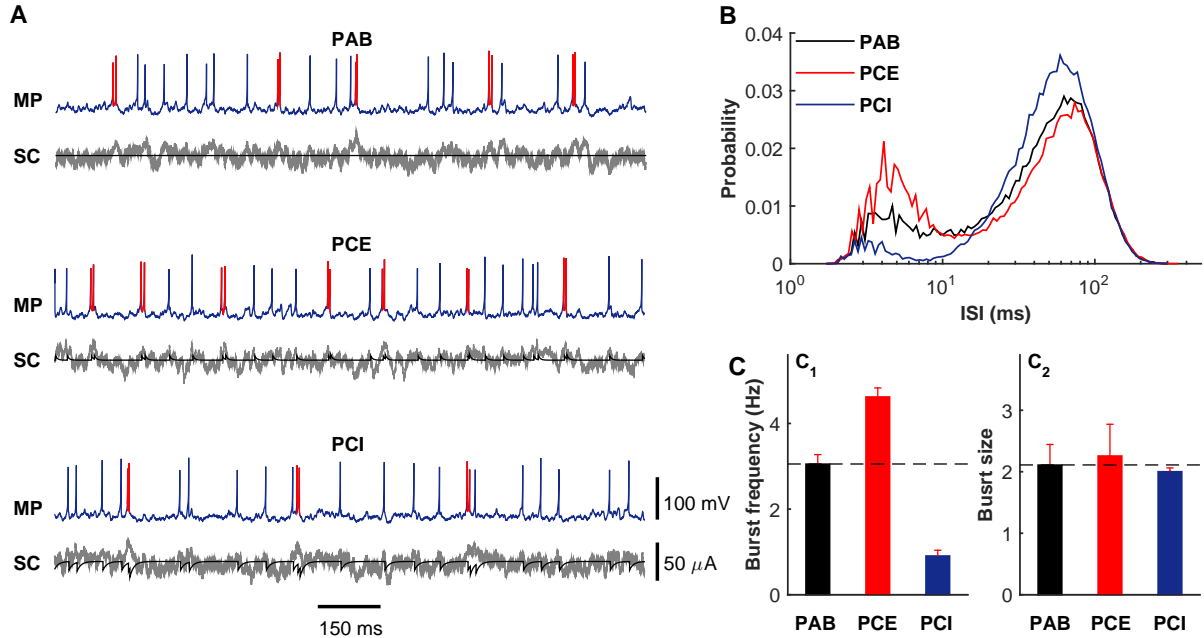}
\caption{(Color online)
Statistics of burst firing explains the mechanisms of chemical autapses in shaping irregular neuronal firing. A:~Typical MP traces and corresponding SC traces for the PAB, PCE and PCI models. Here the red color in MP traces denotes the occurrence of burst firing. The gray lines in SC traces represent the total synaptic currents from presynaptic neurons, and the black lines in SC traces are the autaptic currents from the postsynaptic neuron. B:~ISI distribution curves for the PAB, PCE and PCI models. C:~Burst frequency (C$_1$) and burst size (C$_2$) for the PAB, PCE and PCI models. In simulations, we set $f_{\text{in}}=40$~Hz, and choose $W_{\text{aut}}=0.1$~mS/cm$^2$ for the PCE model and $W_{\text{aut}}=0.6$~mS/cm$^2$ for the PCI model. Note that the black dashed lines in (C$_1$) and (C$_2$) represent the burst frequency and burst size of the PAB model, respectively.}
\label{fig:1}
\end{figure}

To shed insights into why the relatively weak self-feedback input contributed by a chemical autapse can significantly influence the intrinsic proprieties of irregular neuronal firing, we stimulate the postsynaptic neuron using 40~Hz presynaptic trains under different chemical autaptic coupling conditions. Figure 4A illustrates the typical membrane potential traces and corresponding synaptic currents for the PAB, PCE and PCI models, respectively. Compared with the case of autapse blockade, we intuitively observe that the postsynaptic neuron in the PCE model fires more bursts whereas, in contrast, the postsynaptic neuron in the PCI model emits less bursts (Fig.~4A). One possible mechanism to explain this is that the temporally delayed self-excitation and self-inhibition currents tend to boost and reduce the membrane potential of the postsynaptic neuron that just fired, thus providing a biophysical basis to facilitate and suppress burst firing. Indeed, these above observations are also supported by the ISI distributions for different autaptic coupling conditions presented in Fig.~4B, showing that the ISI distribution curves for the PCE and PCI models display the largest and smallest local peaks at small intervals (less than 10~ms). To ascertain what intrinsic properties of burst firing are modulated by chemical autapses, we further perform statistical analysis on the bursting data generated by different models in Fig.~4C. Our results reveal that both excitatory and inhibitory self-feedback inputs primarily regulate the burst frequency (Fig.~4C$_1$) but only slightly change the size of burst firing (Fig.~4C$_2$). However, it should be noted that burst firing of the postsynaptic neuron requires relatively strong presynaptic bombardment, which occasionally provides short-term strong presynaptic current to drive the neuron to emit several high-frequency spikes (see Figs.~2B and 2C). This may explain why irregular neuronal firing modulations caused by chemical autapses mainly observed in the intermediate and high input regimes.

\begin{figure}[!t]
\includegraphics[width=16cm]{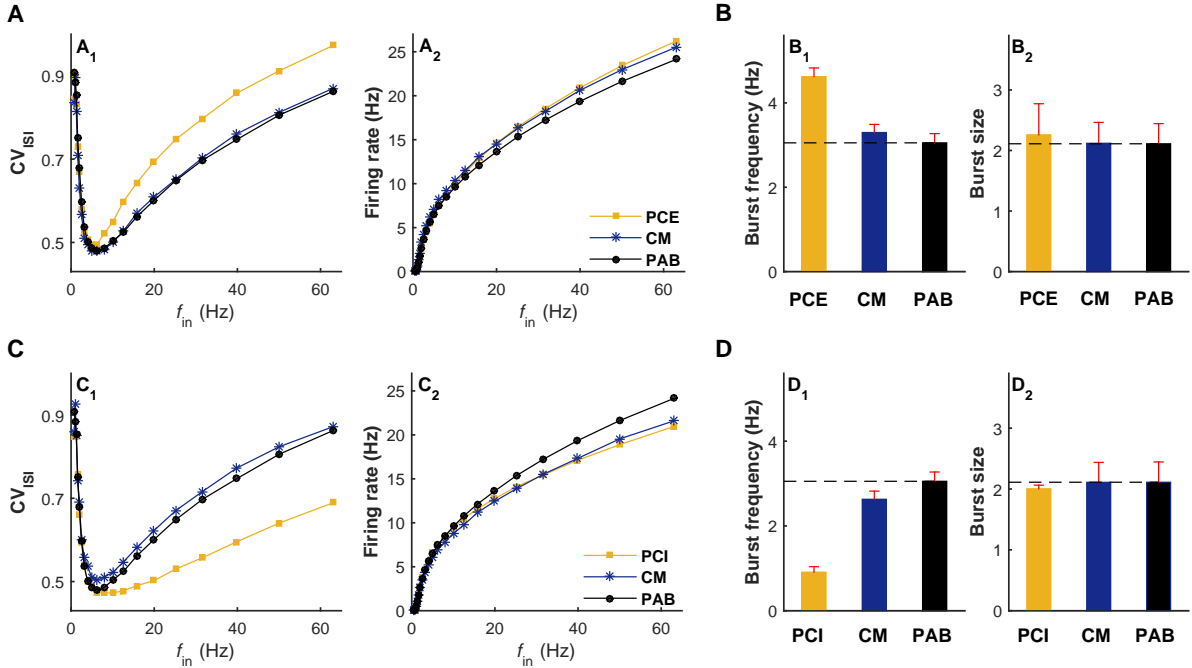}
\caption{(Color online)
Chemical autaptic modulation on irregular neuronal firing cannot be accomplished by an equivalent ``feedforward'' synapse with stochastic inputs. A:~Dpendence of the $\text{CV}_{\text{ISI}}$ value (A$_1$) and average output firing rate (A$_2$) on the input rate $f_{\text{in}}$ for the PCE, CM and PAB models. B:~Burst frequency (B$_1$) and burst size (B$_2$) for the PCE, CM and PAB models, which are driven by $f_{\text{in}}=40$~Hz presynaptic trains. In (A-B), we set $W_{\text{aut}}=0.1$~mS/cm$^2$ for the PCE model and the same coupling strength for the equivalent excitatory ``feedforward'' synapse in the CM model. C:~Dependence of the $\text{CV}_{\text{ISI}}$ value (C$_1$) and average output firing rate (C$_2$) on the input rate $f_{\text{in}}$ for the PCI, CM and PAB models. D:~Burst frequency (D$_1$) and burst size (D$_2$) for the PCI, CM and PAB models, which are driven by $f_{\text{in}}=40$~Hz presynaptic trains. In (C-D), we set $W_{\text{aut}}=0.6$~mS/cm$^2$ for the PCI model and the same coupling strength for the equivalent inhibitory ``feedforward'' synapse in the CM model. It should be noted that the black dashed lines in (B$_1$, D$_1$) and (B$_2$, D$_2$) represent the burst frequency and burst size of the PAB model, respectively.}
\label{fig:1}
\end{figure}

An important question is whether the similar modulations of irregular neuronal firing due to chemical autapses can be also accomplished by normal ``feedforward'' synapses. To address this question, we develop a comparative model (the CM model, Fig.~1B$_4$) for the PCE and PCI models, in which the chemical autapse is replaced by an equivalent Poisson spike train with both the same coupling type and firing rate. Figures~5A and 5C summarize how the $\text{CV}_{\text{ISI}}$ value and the output firing rate are changed with the increasing of input rate for different considered models. By comparing with the results of the PAB model, we observe that stochastic input from an equivalent Poisson spike train in the CM model slightly regulates the average firing rate of postsynaptic neuron (Figs.~5A$_2$ and 5C$_2$) but, as expected, almost does not impact spike-time irregularity (Figs. 5A$_1$ and 5C$_1$). This is not so surprising because, theoretically, the overall contribution of the equivalent Poisson spike train to the total synaptic current still remains at a low level in the CM model (data not shown). Under this condition, the loss of temporal match between the extra input from the equivalent Poisson spike train and membrane potential of postsynaptic neuron considerably weakens the shaping effect on burst firing (Figs.~5B and 5D), and thus does not change the $\text{CV}_{\text{ISI}}$ value significantly compared to the PAB model. Accordingly, we predict that the similar modulations on irregular neuronal firing due to chemical autapses cannot be simply accomplished by normal ``feedforward'' synapses with equivalent but stochastic inputs.

These findings provide the first computational evidence that chemical autapses may shape neuronal firing irregularity in the intermediate and high input regimes. Crucially, we show that the autaptic excitation deteriorates neuronal firing regularity through facilitating bursts, whereas the autaptic inhibition improves neuronal firing regularity by suppressing bursts. Moreover, our results also emphasize the importance of the temporally matching degree between input and output, which might control the modulation capabilities of irregular neuronal firing caused by chemical autapses.

\subsection*{Effect of electrical autapse on modulating irregular neuronal firing}
We next turn to the electrical autaptic coupling and determine how this type of autapse affects the irregular neuronal firing. Signal transmission at an electrical synapse does not require presynaptic action potential and is more efficient than that in chemical synapses~\cite{GerstnerandKistler2002}. In essence, an electrical autapse may contribute not only depolarizing current but also hyperpolarizing current to the postsynaptic neuron at different time instants, which might influence the neuronal dynamics and provide an alternative approach to regulate irregular neuronal firing.

\begin{figure}[!t]
\includegraphics[width=16cm]{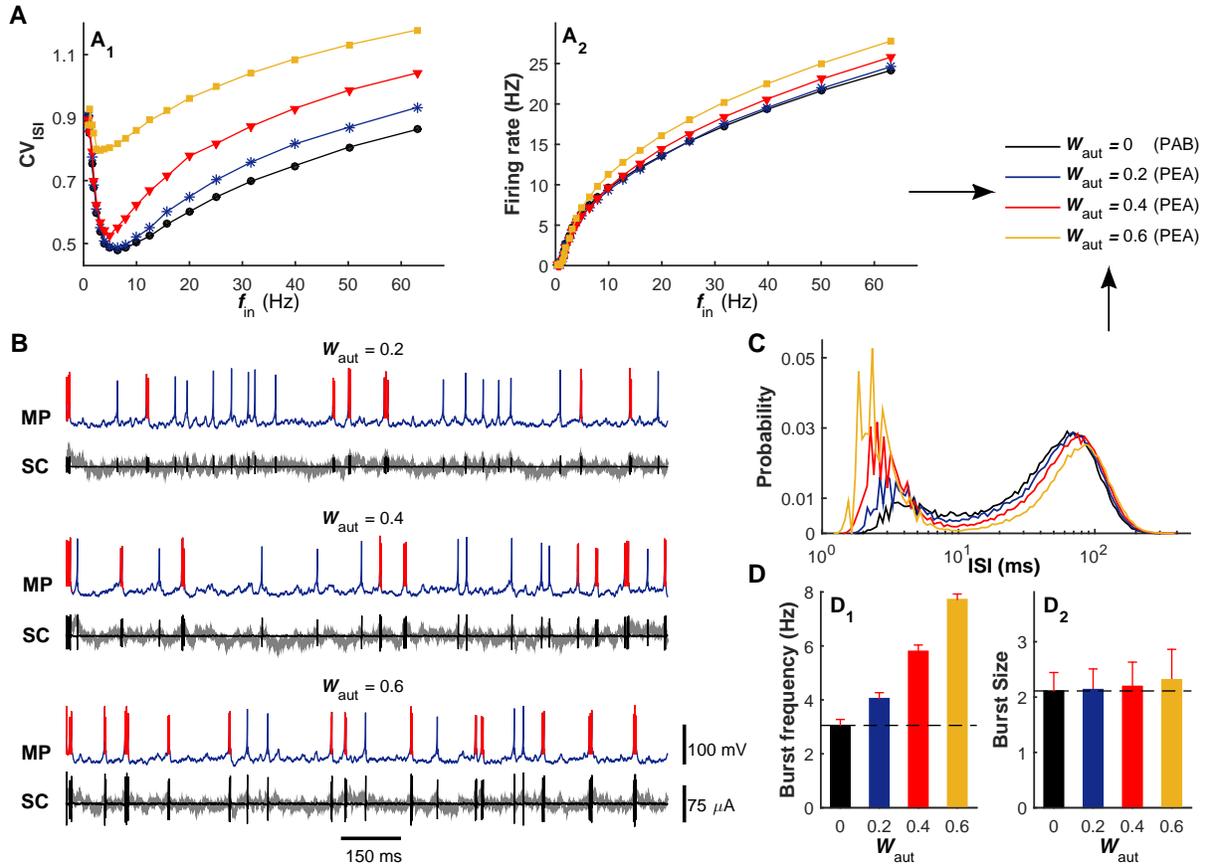}
\caption{(Color online)
Effect of electrical autapse in modulating irregular neuronal firing. A:~The $\text{CV}_{\text{ISI}}$ value (A$_1$) and average output firing rate (A$_2$) are plotted as a function of the input rate $f_{\text{in}}$ for the PEA and PAB models. B:~Typical MP traces and corresponding SC traces of the PEA model at different self-feedback levels. Similar to Fig.~4A, the red color in MP traces denotes the occurrence of burst firing, the gray lines in SC traces represent the total synaptic currents from presynaptic neurons, and the black lines in SC traces are the autaptic currents from the postsynaptic neuron. C:~ISI distribution curves for the PEA and PAB models, with the input rate $f_{\text{in}}=40$~Hz. D:~Burst frequency (D$_1$) and burst size (D$_2$) for the PEA and PAB models. In simulations, electrical coupling strengths are: $W_{\text{aut}}=0$~mS/cm$^2$ (PAB), $W_{\text{aut}}=0.2$~mS/cm$^2$ (PEA), $W_{\text{aut}}=0.4$~mS/cm$^2$ (PEA) and $W_{\text{aut}}=0.6$~mS/cm$^2$ (PEA). The black dashed lines in (D$_1$) and (D$_2$) represent the burst frequency and burst size of the PAB model, respectively.}
\label{fig:1}
\end{figure}

Similar to above studies, we estimate the functional role of electrical autapse by measuring both the $\text{CV}_{\text{ISI}}$ value and output firing rate of the postsynaptic neuron with electrical autapse (the PEA model, Fig.~1B$_5$) for different input rates and comparing them with those generated by the PAB model (Fig.~6A, $W_{\text{aut}}=0$~mS/cm$^2$). At the low self-feedback level, we observe that the pronounced regulation of irregular neuronal firing exists in the intermediate and high input regimes, consistent with the cases of the PCE and PCI models. Under this condition, the weak self-feedback input from the electrical autapse slightly disrupts the neuronal firing regularity and almost unaffects the output firing rate (Figs.~6A and 6B, $W_{\text{aut}}=0.2$~mS/cm$^2$). As the self-feedback level grows, both the depolarizing and hyperpolarizing parts of the autaptic current are enhanced significantly (Fig.~6B, compare the autaptic currents at different self-feedback levels). For a high self-feedback level, the strong depolarizing part of autaptic current drives the postsynaptic neuron to emit burst firing at a relatively high frequency. Note that this strong depolarization causes a drastic modulation on irregular neuronal firing even in the low input regime (Figs.~6A and 6B, $W_{\text{aut}}=0.6$~mS/cm$^2$), which has never been observed for any type of chemical autpase even when the self-feedback level is high (Figs.~3A and 3D). This might be because signal transmission at an electrical autapse is more efficient and stronger than that in chemical autapses, and thus the electrical autaptic current is sufficiently strong to induce and modulate burst firing in the low input regime at a high self-feedback level. However, such strong depolarization does not lead to an unrestricted explosion of high-frequency firing activities, because each bursting event is rapidly terminated by the subsequent and strong hyperpolarizing part of the autaptic current (Fig.~6B). As a consequent, we observe that the burst frequency is enhanced markedly with increasing the electrical autaptic strength (Figs.~6C and 6D$_1$), while the size of burst firing is only slightly improved during this process (Fig. 6D$_2$).

Our results imply that the self-feedback input from electrical autapse destroys neuronal firing regularity through boosting the frequency of burst firing. Compared with the excitatory autapse, the autaptic self-innervation mediated via gap junction exhibits a similar, but more powerful, modulation capability on irregular neuronal firing at the high self-feedback level.

\subsection*{Autaptic transmission delay contributes to the modulation of neuronal firing irregularity}
\begin{figure}[!t]
\includegraphics[width=16cm]{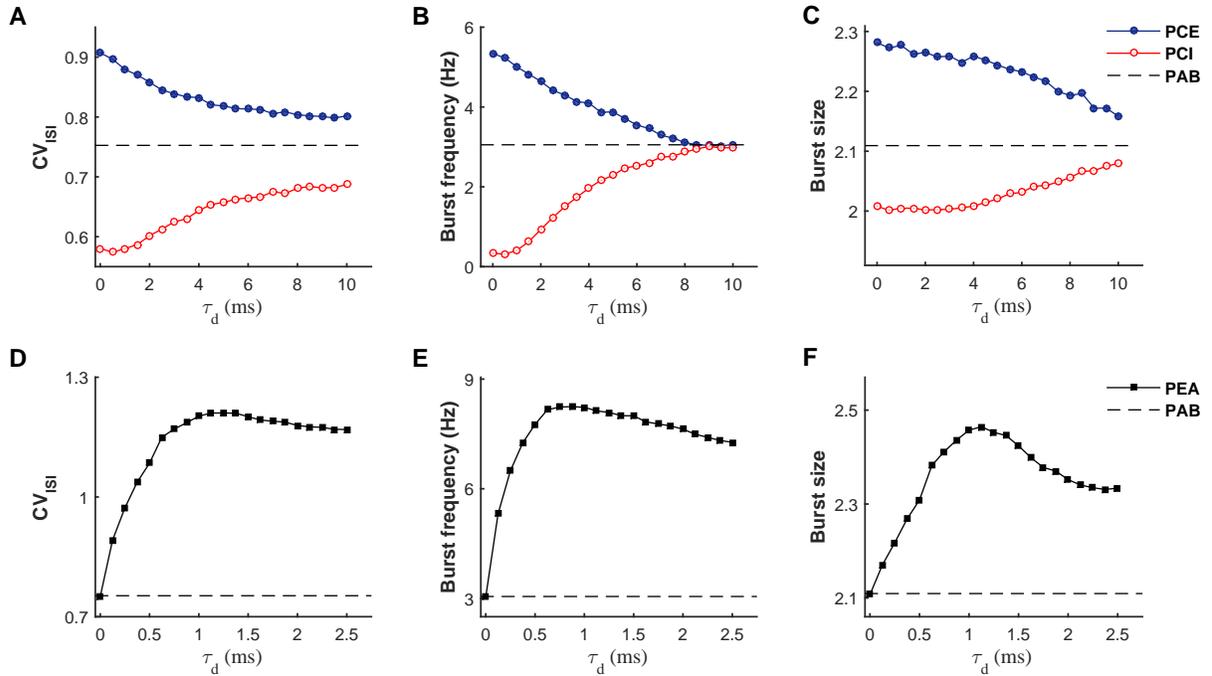}
\caption{(Color online)
Autaptic transmission delay participates into the modulation of irregular neuronal firing. A-C:~Dependence of the $\text{CV}_{\text{ISI}}$ value (A), burst frequency (B) and burst size (C) on the autaptic transmission delay $\tau_{\text{d}}$ for the PCE and PCI models. In each model, the postsynaptic neuron is driven by $f_{\text{in}}=40$~Hz presynaptic trains. In simulations, we set $W_{\text{aut}}=0.1$~mS/cm$^2$ for the PCE model and $W_{\text{aut}}=0.6$~mS/cm$^2$ for the PCI model. The dashed lines in (A-C) represent the default $\text{CV}_{\text{ISI}}$ value, burst frequency and burst size of the PAB model, respectively. D-F:~Dependence of the $\text{CV}_{\text{ISI}}$ value (D), burst frequency (E) and burst size (F) on the autaptic transmission delay $\tau_{\text{d}}$ for the PEA model, which is driven by $f_{\text{in}}=40$~Hz presynaptic trains. In simulations, we choose $W_{\text{aut}}=0.6$~mS/cm$^2$ for the PEA model. As a comparison, the default $\text{CV}_{\text{ISI}}$ value, burst frequency and burst size of the PAB model are also depicted in (D-F), respectively.}
\label{fig:1}
\end{figure}

Because of the finite propagation speed and time lapse occurring by synaptic processing, the synaptic transmission delay is regarded as an important intrinsic property of neural information integration and cannot be neglected~\cite{Kandeletal1991}. There is a rich literature suggesting that the synaptic transmission delay may enrich collective behaviors of neural systems. For instance, it has been reported that introducing an appropriate delay into a neural system may induce synchronization~\cite{MaexandDeSchutter2003}, tame firing patterns~\cite{Masolleretal2008}, facilitate spatio-temporal pattern formation~\cite{Wangetal2009} and modulate burst-type firing~\cite{Guoetal2012}.

To assess whether the transmission delay of chemical autapses also participates into the regulation of irregular neuronal firing, we feed 40~Hz presynaptic trains into the postsynaptic neuron and quantify the $\text{CV}_{\text{ISI}}$ value versus the delay parameter $\tau_{\text{d}}$ for both the PCE and PCI models (see Fig.~7A). As a comparison, the default $\text{CV}_{\text{ISI}}$ value of the PAB model driven by 40~Hz presynaptic trains is also plotted in Fig.~7A (dashed line). For the PCE model, we find that the $\text{CV}_{\text{ISI}}$ curve decays from a high value with the increasing of transmission delay, whereas the opposite trend is observed for the PCI model. Accordingly, these two $\text{CV}_{\text{ISI}}$ curves gradually approach to the dashed line by increasing the self-feedback transmission delay (Fig.~7A). These observations provide the evidence that both excitatory and inhibitory autapses require relatively fast spike transmission speed to ensure their strong modulation capabilities on irregular neuronal firing. Mechanistically, this might be because slow self-feedback inputs from chemical autapses tend to delay the membrane potential response of the postsynaptic neuron that just fired, thus reducing their shaping effects on burst firing. To validate whether this is correct, we calculate both the frequency and size of burst firing as a function of the transmission delay. Our results presented in Figs.~7B and 7C confirm that a long transmission delay indeed significantly weakens the shaping effects of chemical autpases on both the frequency and size of burst firing.

We then perform the similar analysis for the PEA model to determine how the transmission delay of electrical autapse affects neuronal firing irregularity. Because electrical synapse conduces signal much faster than chemical synapses, a relatively small time delay interval is considered for the electrical autapse in our study. As the transmission delay is increased, we observe that the $\text{CV}_{\text{ISI}}$ curve first rises rapidly and then declines slowly, and the largest $\text{CV}_{\text{ISI}}$ value is achieved at an intermediate transmission delay (Fig.~7D). This noticeable feature of the $\text{CV}_{\text{ISI}}$ curve indicates that the electrical autapse exhibits the strongest modulation capability on irregular neuronal firing at an optimal transmission delay. Further statistical analysis shows that both the frequency and size of burst firing capture a similar pattern as the $\text{CV}_{\text{ISI}}$ curve (Figs.~7E and 7F), suggesting that the depolarizing capability of the self-feedback input is strong at intermediate transmission delays. This is reasonable because, due to the oscillatory feature of membrane potential, an intermediate transmission delay of electrical autapse can guarantee relatively large voltage difference for a short, but sufficient, period of time just after the postsynaptic neuron spiking, which provides a biophysical basis to trigger burst firing at a higher frequency.

These observations highlight the importance of autaptic transmission delay in mediating the neuronal firing regularity. Interestingly, we show that chemical autapes require a short transmission delay to ensure strong modulations on irregular neuronal firing, whereas the electrical autapse may achieve its strongest modulation capability at an optimal transmission delay. As a regulation parameter, the existence of autaptic transmission delay enriches the variability of neuronal firing, and tuning its value enables spiking irregularity of neurons to vary in a certain range.

\subsection*{Our results can be extended to spiking neurons with class II and III excitabilities}
So far, we have shown that different types of autapses may shape the irregular firing of class I neuron in different manners. In addition to the class I neuron, neurons with class II and III excitabilities are also ubiquitous in the brain~\cite{Prescottetal2008}. Thus, a naturally arising question is whether the similar modulations on irregular neuronal firing due to different types of autapses can be also observed for neurons exhibiting class II and III excitabilities. We try to answer this question by performing additional simulations, using two novel sets of model parameters~\cite{Izhikevich2003}: (1) $a=0.02$, $b=0.2$, $c=-65$~mV and $d=2$; (2) $a=0.02$, $b=0.25$, $c=-65$~mV and $d=6$. With these two choices, the postsynaptic neuron in our model displays typical class II and III excitabilities.

\begin{figure}[!t]
\includegraphics[width=16cm]{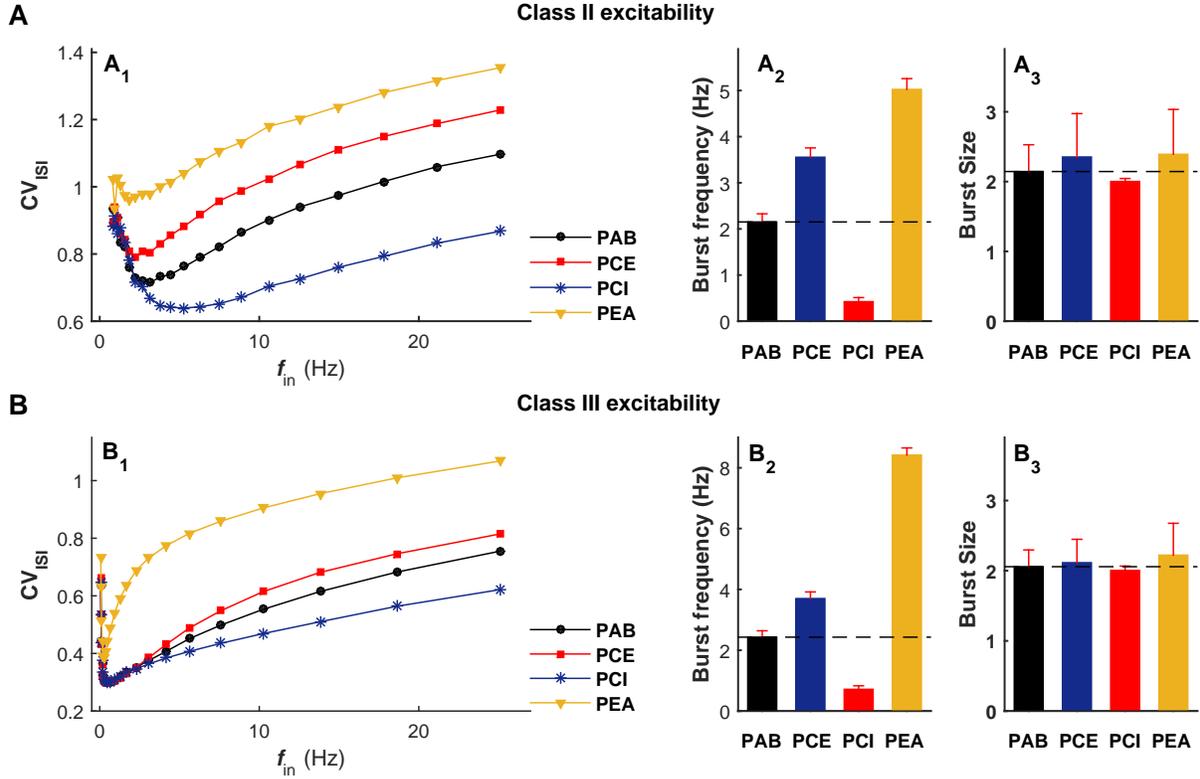}
\caption{(Color online)
The main results based on class I neuron are extendable to spiking neurons with class II and III excitabilities. A, B: ~Simulations of the postsynaptic neuron with class II excitability (A) and with class III excitability (B). In (A$_1$, B$_1$), the $\text{CV}_{\text{ISI}}$ value is plotted as a function of the input rate $f_{\text{in}}$ under different autaptic coupling conditions. In (A$_2$, A$_3$), we show the burst frequency and burst size of the class II postsynaptic neuron driven by $f_{\text{in}}=8$~Hz presynaptic trains. Similarly, the burst frequency and burst size of the class III postsynaptic neuron driven by $f_{\text{in}}=16$~Hz presynaptic trains are illustrated in (B$_2$, B$_3$). For the class II postsynaptic neuron, we set $W_{\text{aut}}=0.05$~mS/cm$^2$ for the PCE model, $W_{\text{aut}}=0.3$~mS/cm$^2$ for the PCI model and $W_{\text{aut}}=0.3$~mS/cm$^2$ for the PEA model. For the class III postsynaptic neuron, we set $W_{\text{aut}}=0.05$~mS/cm$^2$ for the PCE model, $W_{\text{aut}}=0.3$~mS/cm$^2$ for the PCI model and $W_{\text{aut}}=0.5$~mS/cm$^2$ for the PEA model.
}
\label{fig:1}
\end{figure}

In Figs.~8A and 8B, we illustrate our detailed simulation results for neurons exhibiting class II and III excitabilities, respectively. There are at least three main observations are consistent with our above key findings for the class I neuron, which are summarized as follows. First, both excitatory and inhibitory autapses are found to play important roles in mediating irregular neuronal firing, but in different modulation manners (Figs.~8A$_1$ and 8B$_1$). More specifically, the autaptic excitation reduces firing regularity due to burst enhancement, whereas the autaptic inhibition improves the firing regularity because of burst suppression (Figs.~8A$_2$ and 8B$_2$). Second, the electrical autapse might noticeably promote burst firing (Figs.~8A$_2$ and 8B$_2$), thus seriously worsening neuronal firing irregularity. By comparing with the results of PEC and PCI models, it is observed that the autaptic self-innervation mediated via gap junction has a stronger modulation capability within a wider input rate regime (Figs.~8A$_1$ and 8B$_1$). Third, self-feedback inputs from different types of autapses mainly regulate the frequency of burst firing (Figs.~8A$_2$ and 8B$_2$) and affect limited to burst size (Figs.~8A$_3$ and 8B$_3$).

These results suggest that our key findings based on the class I neuron are not qualitatively impacted by single neuron firing properties and can be extendable to spiking neurons with class II and III excitabilities, further emphasizing the generality and functional importance of autaptic transmission in modulating irregular neuronal firing.

\section*{Discussion}
Cortical neurons have been observed to discharge highly irregular, seemingly random, action potentials in vivo~\cite{Stiefeletal2013, Fellousetal2003, Destexheetal2003}. There is accumulating experimental evidence that the temporally irregular firing of neurons might be tightly associated with several higher brain cognitive functions~\cite{HanselandMato2013, Ardidetal2010, Doronetal2014}. Understanding the origin and modulation of irregular neuronal firing is therefore an important topic in computational neuroscience. Although several past modelling studies have established that irregular firing of neurons might arise from strongly balanced excitation-inhibition presynaptic bombardment~\cite{Destexheetal2003, PikovskyandKurths1997, Fellousetal2003, Kreuzetal2006, Lucciolietal2006}, so far it still lacks an insightful mechanistic understanding on how different types of autaptic transmission contribute to the regulation of irregular neuronal firing.

Using a biophysical model neuron that incorporates autaptic coupling, we extended previous works and presented the first computational investigation on how chemical and electrical autapses modulate the irregular neuronal firing in this study. Mechanistically, we identified that both excitatory and electrical autapses destroy neuronal firing regularity by boosting the burst frequency, whereas the inhibitory autapse subserves neuronal firing regularity due to burst suppression. Compared with chemical autapses, we found that the electrical autapse exhibits a stronger modulation capability on irregular neuronal firing, which can be observed even at the low level of presynaptic bombardment. Theoretically, this might be because signal transmission at an electrical autapse is more efficient and stronger than that in chemical autapses over a short period after the postsynaptic neuron spiking. Further investigation demonstrated that the autaptic transmission delay also contributes to the modulation of irregular neuronal firing, and tuning its value enables spike-time irregularity to vary in a certain range. Together, these results make an important addition to past studies and stress the underlying functional importance of autaptic transmission in modulating irregular neuronal firing.

We note that parts of these findings are in agreement with former experimental observations, and our results established more insightful mechanistic understandings. For instance, previous electrophysiological recordings have shown that the firing of fast-spiking interneurons from sensorimotor cortex become more irregular when autaptic transmission is blocked by gabazine and can be restored by applying the dynamic-clamp autaptic conductance~\cite{BacciandHuguenard2006}. Similar experiments were performed for pyramidal neurons from the same brain region, revealing that this neuron type also fire more regularly in the presence of artificial GABAergic autaptic transmission~\cite{BacciandHuguenard2006}. These observations provide direct evidence to support our theoretical explanation on how inhibitory autapse shapes irregular neuronal firing. Moreover, a cell culture experiment has found that the excitatory autapses are essential for spontaneous burst firing generated by hippocampal pyramidal neurons~\cite{Wyartetal2005}. To a certain degree, such finding might be related to the autaptic excitation-induced burst enhancement presented in this study, thus indirectly supporting the proposed significant contribution of autaptic excitation to irregular neuronal firing. Nevertheless, so far there still lacks experimental evidence to endorse the functional role of electrical autapse in modulating neuronal firing irregularity, which is of importance and deserves to be clarified in future electrophysiological studies.

Although autapses have been treated as a trivial type of synapses in many studies, increasing anatomical data indicate that self-connected neurons are widely distributed in the brain. By injecting neurons with intracellular markers, autapses have been discovered in multiple brain regions, including the striatum~\cite{Parketal1980}, substantia nigra~\cite{KarabelasandPurpura1980}, hippocampus~\cite{Cobbetal1997} and neocortex \cite{Baccietal2003, Tamasetal1997}. Under different conditions, neurons from these brain regions have been reported to display varieties of operating models. A typical example is the pyramidal neurons stemming from hippocampus and neocortex. Using whole-cell patch-clamp on brain slices, previous studies suggested that regular spiking pyramidal neurons primarily work as integrators~\cite{Tatenoetal2004}. This opinion, however, has been challenged by recent experimental and computational investigations, showing that both shunting inhibition and adaptation greatly mediate the firing proprieties of pyramidal neurons and may make them behave more like resonators or coincidence detectors~\cite{Prescottetal2008}. Furthermore, fast-spiking interneurons were also showed to preferentially work as resonators but~\cite{Tatenoetal2004}, sometimes, they operated as integrators~\cite{Hoetal2012}. Notably, although most simulations presented in this study were performed using the class I neuron, we have further tested and demonstrated that our critical findings can be extendable to neurons with other types of excitabilities. This provides us a strong computational evidence that our proposed functional roles in regulating irregular neuronal firing for different types of autapses are independent of signal neuron firing proprieties, and thus might be regarded as fundamental principles of autaptic modulations on neuronal firing irregularity.

There are several important physiological implications emerging from our current results. On the one hand, our model predicts that the autaptic transmission might be involved in both the generation and regulation mechanisms of burst firing~\cite{BacciandHuguenard2006, Wyartetal2005}. In the brain, many types of neurons have been identified to fire bursts. Due to the existence of synaptic failure, burst firing can guarantee a synapse conducting signal in a reliable manner~\cite{Lisman1997}. We therefore postulate that autaptic self-innervation of neurons might offer underlying biological plausibility to achieve reliable neuronal information transmission~\cite{Yamaguchietal1997}. On the other hand, the irregular neuronal firing modulation induced by autaptic transmission may be associated with several higher brain functions, such as working memory. Past experimental evidence have shown that irregular persistent activity recorded in neocortex is the typical feature of working memory~\cite{HanselandMato2013}. Based on our findings presented in this study, both excitatory and electrical autapses may serve as underlying physiological substrates of highly irregular persistent firing activity during working memory tasks. Additionally, the autaptic transmission might also participate in the control of some brain disorders. For example, the substantia nigra pars reticulata (SNr) is the main output structure of basal ganglia and has been recently found to control absence seizures in a bidirectional manner~\cite{Chenetal2014, Chenetal2015}. Thus, the anatomical identification of autapses of the inhibitory SNr neurons indicates inhibitory autaptic transmission of SNr neurons might contribute to the bidirectional control of absence seizures by the basal ganglia~\cite{KarabelasandPurpura1980}.

Our model is parsimonious, designed to capture basic modulation mechanisms of irregular neuronal firing by autapses, and can be extended in several ways. First, we utilized a fixed threshold detection strategy in this study. However, experimental data showed that biological neurons emit spikes in a variable-threshold fashion~\cite{Fontaineetal2014, WilentandContreras2005}. Introducing an adaptive threshold mechanism into our model can directly impact the firing of postsynaptic neuron, thus affecting the dynamics of autaptic transmission. It will be important to further probe how the spike-threshold adaptation contributes to the modulation of neuronal firing irregularity. Second, we did not consider any plasticity mechanism for chemical autapses. As is well known, synaptic transmission can be enhanced and depressed by pre- and postsynaptic firing activities, with a temporal span ranging from milliseconds to several days~\cite{Tsukahara1981}. Theoretically, autaptic enhancement can enlarge the modulations of chemical synapses on irregular neuronal firing and, in contrast, autaptic depression tends to weaken these shaping effects. In future studies, we need to perform computational studies by integrating synaptic plasticity in our model to test whether our above predictions capture the real fact. Third, cortical neurons are commonly driven by correlated presynaptic inputs. Previous studies have confirmed that neural correlation might play important roles in both modulating neurodynamics~\cite{Kreuzetal2006, Lucciolietal2006} and tuning systemic properties of neural networks, such as memory and patterns formation~\cite{Hopfield1982, Amit1992, Sollich2014, Agliari2012, Agliari2015}. It is interesting and necessary to further investigate how the neural correlation shapes the regulation of irregular neuronal firing by autaptic transmission. Finally, our current model only includes one ``self-feedback'' autapse, but many experimental studies have suggested that a real biological neuron might contain multiple autapses~\cite{Lubkeetal1996, Tamasetal1997, Cobbetal1997}. Future work should explore the complicated combination roles of multiple autapses in the modulation of irregular neuronal firing.

In conclusion, we have systematically performed mechanistic studies to investigate the functional importance of autaptic transmission in modulating neuronal firing irregularity. Our results showed that both excitatory and electrical autapses subserve the production of burst firing, thus contributing negatively to neuronal firing regularity. Contrarily, the inhibitory autapse was found to suppress burst firing and therefore improve neuronal firing regularity. These observations established an insightful mechanistic understanding on how different types of autapses shape the irregular firing at the single-neuron level. Notably, we disclosed that these proposed autaptic modulation mechanisms on irregular neuronal firing are general principles, which are applicable to neurons operating in different modes. Recent fast developments in high-resolution dynamic-clamp and voltage-clamp recording techniques could be utilized to validate testable predictions offered by our model. Finally, we note that as a promising tool the phase response curve can be used to further characterize the complicated dynamical responses of neurons with different types of autapses due to perturbations in future studies~\cite{Levnajic2010, Perez2007}.

\section*{Materials and Methods}

\noindent

\subsection*{Computational model}
We consider a single postsynaptic neuron receiving both the ``feedforward'' balanced excitation-inhibition inputs from totally $N$ presynaptic neurons and the ``self-feedback'' autaptic input from itself. As schematically shown in Fig.~1A, we choose $N_{\text{ex}}=\rho N$ presynaptic neurons as excitatory and the rest $N_{\text{inh}}=(1-\rho) N$ presynaptic neurons as inhibitory. Unless otherwise stated, we set $N=1000$ and choose $\rho=0.8$ in our model as the ratio of excitatory to inhibitory neurons is reported to be approximate to 4:1 in mammalian neocortex~\cite{GerstnerandKistler2002}. The firing dynamics of the postsynaptic neuron is simulated by using the simple model neuron proposed recently by Izhikevich~\cite{Izhikevich2003}. The subthreshold membrane potential of the Izhikevich model neuron obeys the following two differential equations~\cite{Izhikevich2003}:
\begin{equation}
\begin{split}
\frac{dv}{dt} = 0.04v^2 + 5v+ 140 -u + \frac{1}{C}({I_{\text{pre}} + I_{\text{aut}}}),
\end{split}
\label{eq:1}
\end{equation}
\begin{equation}
\begin{split}
\frac{du}{dt} = a(bv-u),
\end{split}
\label{eq:2}
\end{equation}
where $C=1$~$\mu$F/cm$^2$, $v$ represents the membrane potential (in~mV), $u$ denotes the slow membrane recovery variable, $I_{\text{pre}}$  is the total synaptic current receiving from presynaptic neurons and $I_{\text{aut}}$ is the autaptic current, respectively. Unless otherwise specified, four model parameters have the following values: $a=0.02$, $b=0.2$, $c=-65$~mV and $d=8$. This choice corresponds to a neuron working as integrator and exhibiting class I excitability~\cite{Izhikevich2003}. Note that the main results shown in this study are obtained using the class I neuron. In additional simulations, we demonstrate the similar results can be also observed for class II (resonator) and class III (coincidence detector) neurons~\cite{Prescottetal2008}. For the Izhikevich model neuron, a spike is detected whenever the potential reaches the peak of the spike ( $v_{\text{peak}}=30$~mV), and then the membrane potential and recovery variable are reset according to: $v \leftarrow c$ and $u \leftarrow u+d$.

In our study, each presynaptic neuron is assumed to be a simple spike generator and emits spikes in an independent Poisson fashion with the same input rate $f_{\text{in}}$  (in~Hz). We model the total synaptic current receiving from ``feedforward'' presynaptic neurons as current-based:
\begin{equation}
\begin{split}
I_{\text{pre}} = G_{\text{ex}}(E_{\text{ex}}-V_{\text{rest}}) + G_{\text{inh}}(E_{\text{inh}}-V_{\text{rest}}).
\end{split}
\label{eq:3}
\end{equation}
Here $G_{\text{ex}}$ and $G_{\text{inh}}$ are the total excitatory and inhibitory presynaptic conductances, $E_{\text{ex}}=0$~mV and $E_{\text{inh}}=-80$~mV are reversal potentials for excitatory and inhibitory synapses, and $V_{\text{rest}}=-60$~mV is the resting membrane potential, respectively. When a presynaptic neuron emits a spike, a fixed increment is assigned to corresponding presynaptic conductance: $G_{\text{ex}} \leftarrow G_{\text{ex}}+W_{\text{ex}}$ for an excitatory spike and $G_{\text{inh}} \leftarrow G_{\text{inh}}+W_{\text{inh}}$ for an inhibitory spike. Otherwise, these two parameters decay exponentially as follows:
\begin{equation}
\begin{split}
\tau_{\text{ex}}\frac{dG_{\text{ex}}}{dt} = -G_{\text{ex}}
\end{split}
\label{eq:4}
\end{equation}
and
\begin{equation}
\begin{split}
\tau_{\text{inh}}\frac{dG_{\text{inh}}}{dt} = -G_{\text{inh}},
\end{split}
\label{eq:5}
\end{equation}
with synaptic time constants $\tau_{\text{ex}}=5$~ms and $\tau_{\text{inh}}=10$~ms. Note that two synaptic variables $W_{\text{ex}}$ and $W_{\text{inh}}$ are relative peak conductances of excitatory and inhibitory presynapses that determine their coupling strengths. In the following simulations, we fix the value of $W_{\text{ex}}$  and set
\begin{equation}
\begin{split}
W_{\text{inh}}=\frac{(E_{\text{ex}}-V_{\text{rest}})\cdot N_{\text{ex}}\cdot \tau_{\text{ex}}}{(E_{\text{inh}}-V_{\text{rest}})\cdot N_{\text{inh}}\cdot \tau_{\text{inh}}}W_{\text{ex}}.
\end{split}
\label{eq:6}
\end{equation}
Under this condition, the received excitation and inhibition of the postsynaptic neuron from its presynaptic neurons are theoretically perfect balanced. Unless otherwise specified, we choose $W_{\text{ex}}=0.01$~mS/cm$^2$ in this work.

In the present study, we consider both chemical and electrical autaptic connections. Depending on the type of postsynaptic neuron, the chemical autapse is either excitatory or inhibitory in our model. Similar to the total presynaptic current, the ``self-feedback'' chemical autaptic current is simply modelled as follows:
\begin{equation}
\begin{split}
I_{\text{aut}}=G_{\text{aut}}(E_{\text{aut}}-V_{\text{rest}}),
\end{split}
\label{eq:7}
\end{equation}
where $G_{\text{aut}}$ is the autaptic conductance and $E_{\text{aut}}$ is the reversal potential ($E_{\text{aut}}=0$~mV for excitatory autapse and $E_{\text{aut}}=-80$~mV for inhibitory autapse). Whenever the postsynaptic neuron emits a spike, the autaptic conductance is increased after a fixed transmission delay $\tau_{\text{d}}$ according to: $G_{\text{aut}} \leftarrow G_{\text{aut}} +W_{\text{aut}}$, where parameter $W_{\text{aut}}$ represents the autaptic coupling strength. In our studies, we chose $W_{\text{aut}}=h\cdot W_{\text{ex}}$ for excitatory autapse and $W_{\text{aut}}=h\cdot W_{\text{inh}}$ for inhibitory autapse, respectively. Otherwise, the chemical autaptic conductance decays exponentially with a fixed time constant ($\tau_{\text{aut}}=5$~ms for excitatory autapse and $\tau_{\text{aut}}=10$~ms for inhibitory autapse). Moreover, the autapse might be mediated via gap junction (electrical autapse) for inhibitory postsynaptic neuron, modelled as follows:
\begin{equation}
\begin{split}
I_{\text{aut}}=W_{\text{aut}}(v(t-\tau_{\text{d}})-v),
\end{split}
\label{eq:8}
\end{equation}
where $W_{\text{aut}}$  is the autaptic coupling strength and  $\tau_{\text{d}}$ is a fixed transmission delay. Unless otherwise stated, we set $\tau_{\text{d}}=2$~ms for chemical (excitatory and inhibitory) autapses and $\tau_{\text{d}}=0.5$~ms for electrical autapse in the following simulations.

\subsection*{Data analysis}
We employ several data analysis techniques to quantitatively evaluate the spike trains generated by our model. To characterize the temporal regularity of spike trains, the coefficient of variation (CV) of inter-spike intervals (ISIs) is utilized. Mathematically, the coefficient of variation of ISIs is defined as~\cite{KochandSegev1998}:
\begin{equation}
\begin{split}
{\text{CV}}_{\text{ISI}}=\frac{\sqrt{\langle T_{i}^2\rangle - \langle T_{i}\rangle^2}}{\langle T_i \rangle},
\end{split}
\label{eq:9}
\end{equation}
where the symbol $\langle\cdot\rangle$ denotes the average over time, $T_i = t_{i+1}-t_i$, and $t_i$ is the time of the $i$-th firing of the postsynaptic neuron. By definition, increased ${\text{CV}}_{\text{ISI}}$ reflects an increased interval-to-interval variability and thus a decreased regularity of neuronal firing. Throughout our studies, the reported ${\text{CV}}_{\text{ISI}}$ values are averaged over 50 independent trials with different random seeds.

In some cases, we estimate the probability distribution curve of ISIs. For each experimental setting, the ISI distribution curve is obtained based on 10$^5$ firing events. In this study, the burst firing is defined as a groups of spikes (at least two spikes) with intervals between two successive spikes less than 10~ms. Statistically, we note that burst firing can be well identified provided that a short and noticeable ISI peak appears in the ISI distribution curve. To further evaluate the intrinsic properties of burst firing, we perform further statistical analysis on both burst frequency and size for several recording data. The burst frequency is computed as the average number of burst firing per second based on 120 trials with each trial of 50 seconds simulation, and the burst size is calculated as the average number of spikes contained in a burst event using all bursting data generated above.

To evaluate the contribution of self-feedback input to the total synaptic current, we compute the contribution factors for both the excitatory and inhibitory autaptic currents in this study. As a dimensionless measure, the contribution factor (CF) is calculated as follows:
\begin{equation}
\begin{split}
\text{CF}=\frac{f_{\text{out}}\cdot h}{f_{\text{in}}\cdot (N_{\text{ex}} + N_{\text{inh}})},
\end{split}
\label{eq:8}
\end{equation}
where $f_{\text{out}}$ is the average output firing rate of the postsynaptic neuron. A larger CF value corresponds to a greater contribution of self-feedback input to the total synaptic current.

\subsection*{Simulation details}
Model simulations and data analysis are performed in MATLAB environment (MathWorks, USA). The differential equations described above are intergraded numerically using the Euler algorithm with a fixed time step of $dt=0.1$~ms. The chosen integration time step is demonstrated to be small enough to ensure an accurate simulation of the Izhikevich model neuron. For each simulation, the initial membrane potential of the postsynaptic neuron is uniformly distributed between -70 and 30~mV, and the slow membrane recovery variable is initially set as $u=bv$~\cite{Izhikevich2003}. Note that we carry out all simulations for sufficiently long time (at least 50 seconds) to collect data for further statistical analysis. Computer codes implementing our model will be available from the authors upon request or the Yale Model Database (\url{https://senselab.med.yale.edu/modeldb/}).

\section*{Acknowledgements}
This work is supported by the National Natural Science Foundation of China (No.~81571770, 61527815, 61201278, 81371636, 81401484 and 81330032).

\section*{Author Contributions}
D.G. S.W. and D.Y. designed experiments,  D.G. S.W. J.M. Y.C. and M.C. performed experiments, D.G. S.W. Y.Z. J.M. M.C. P.X. Y.X. D.Y. and M.P. analyzed the data, and D.G. M.P. D.Y. and Y.Z. wrote the manuscript.

\section*{Competing Financial Interests}
The authors declare no competing financial interests.

\end{document}